\documentclass[twocolumn, superscriptaddress, prl, amsmath, amssymb, aps, 10pt, floatfix]{revtex4-2}

\usepackage{float}
\usepackage{graphicx}
\usepackage{dcolumn}
\usepackage{bm}
\usepackage{xcolor}
\usepackage{textgreek}
\usepackage{textcomp}
\usepackage[separate-uncertainty=true, multi-part-units=single, list-units=repeat]{siunitx} 
\usepackage[version=4]{mhchem} 
\usepackage[thinc]{esdiff} 
\usepackage{comment}
\usepackage{textcomp}

\begin{document}

\preprint{APS/123-QED}

\title{Using optical tweezers to simultaneously trap, charge and measure the charge of a microparticle in air}

\author{Andrea Stoellner}
\affiliation{Institute of Science and Technology Austria, Am Campus 1, 3400 Klosterneuburg, Austria}
\author{Isaac C.D. Lenton}
\affiliation{Institute of Science and Technology Austria, Am Campus 1, 3400 Klosterneuburg, Austria}
\author{Artem G. Volosniev}
\affiliation{Center for Complex Quantum Systems, Department of Physics and Astronomy, Aarhus University, Ny Munkegade 120, DK-8000 Aarhus C, Denmark}
\author{James Millen}
\affiliation{King's College London, Strand Campus, London WC2R 2LS, United Kingdom}
\author{Renjiro Shibuya}
\affiliation{Chiba University, 1-33, Yayoi-cho, Inage-ku, Chiba 263-8522 Japan}
\author{Hisao Ishii}
\affiliation{Chiba University, 1-33, Yayoi-cho, Inage-ku, Chiba 263-8522 Japan}
\author{Dmytro Rak}
\affiliation{Institute of Science and Technology Austria, Am Campus 1, 3400 Klosterneuburg, Austria}
\affiliation{Institute of Experimental Physics, Slovak Academy of Sciences, Watsonova 47, 040 01 Košice, Slovakia}
\author{Zhanybek Alpichshev}
\affiliation{Institute of Science and Technology Austria, Am Campus 1, 3400 Klosterneuburg, Austria}
\author{Grégory David}
\affiliation{ETH Zürich, Rämistrasse 101, 8092 Zürich, Switzerland}
\author{Ruth Signorell}
\affiliation{ETH Zürich, Rämistrasse 101, 8092 Zürich, Switzerland}
\author{Caroline Muller}
\affiliation{Institute of Science and Technology Austria, Am Campus 1, 3400 Klosterneuburg, Austria}
\author{Scott Waitukaitis}
\email{scott.waitukaitis@ista.ac.at}
\affiliation{Institute of Science and Technology Austria, Am Campus 1, 3400 Klosterneuburg, Austria}

\date{\today}

\begin{abstract}

Optical tweezers are widely used as a highly sensitive tool to measure forces on micron-scale particles. One such application is the measurement of the electric charge of a particle, which can be done with high precision in liquids, air, or vacuum. We experimentally investigate how the trapping laser itself can electrically charge such a particle, in our case a $\sim$1 \textmu{m} SiO$_2$ sphere in air. We model the charging mechanism as a two-photon process which reproduces the experimental data with high fidelity. 

\end{abstract}

\maketitle
Optical tweezers are exquisitely well-suited for sensing forces in settings ranging from the interactions of colloidal particles \cite{Merrill2009, Mitchem2008}, to the chemistry and morphology of microdroplets \cite{Mohajer2025, Sullivan2020}, to precise manipulation of micron-scale objects \cite{Ruffner2012, Rubinsztein-Dunlop2015}. When applied to probe electrical forces, this remarkable sensitivity can be used to measure the charge of small objects, akin to the famous Millikan oil-drop experiment \cite{Millikan1913} with sub-electron precision. This capability has been useful in a variety of areas, including studies aimed at the charging dynamics of colloidal particles in non-polar liquids \cite{Beunis2010, Beunis2012, Schreuer2018, Schreuer2018-2}, the precise determination of particle mass \cite{Ricci2019}, and the measurement of electric fields \cite{Zhu2023}, among others \cite{Pesce2015, Marmolejo2021}. Well before these recent works, the first experiments to measure the charge of an optically levitated particle were carried out by Arthur Ashkin \cite{Ashkin1976}, who observed that the trapping light could also \textit{cause} certain particles to become charged. To the best of our knowledge, however, this effect has not been explored further since.

\begin{figure}[ht!]
\centering
\includegraphics{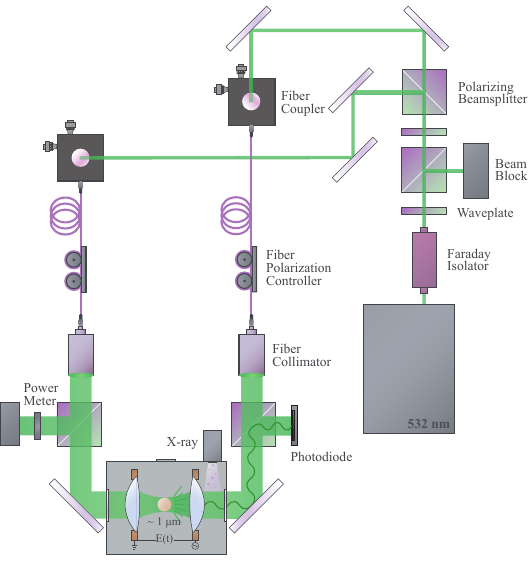}
\caption{ \label{fig:experimental_setup} Experimental setup. The trapping laser (532 nm) is split into two beams by a polarizing beam splitter and coupled into single-mode optical fibers. After being brought back to free space, the beams enter the grounded experimental chamber through one inch windows on either side. Inside, they are focused by two lenses with a focal length of 8~mm to form the trap. Each lens is surrounded by a copper ring electrode. The electric field between these induces oscillations of the particle along the beam axis. We measure the particle's position via changes in the scattered light collected on the photodiode on the right side of the chamber. For particle loading we spray in the aerosol from the port on top of the chamber.}
\end{figure}

\begin{figure}[ht!]
\centering
\includegraphics{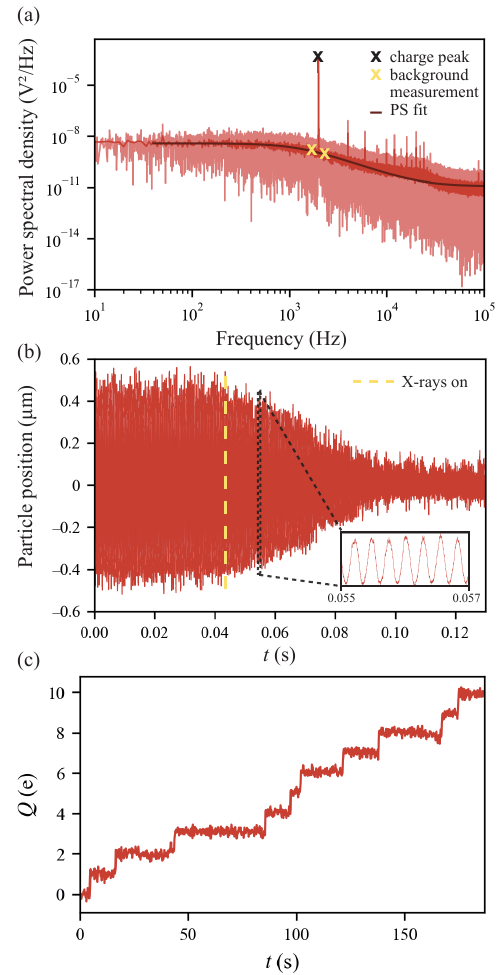}
\caption{ \label{fig:measurement_principle} Measurement principle. (a) The power spectral density (PSD) of the electrically driven particle shows a clear peak at the driving frequency $f_{dr}$ = 2 kHz. To obtain the particle's charge, we perform a lock-in style measurement of this charge peak as well as the background PSD values at $f_{dr} \pm \text{300 Hz}$. (b) The particle position oscillates around the trap center with a frequency $f_{dr}$. After the x-ray source is turned on to discharge the particle, the oscillations decay exponentially. (c) Time evolution of the particle charge showing individual elementary charge steps. 
}
\end{figure}

In this work, we study in detail the non-linear photoelectric charging of a sub-micron particle held in a high-intensity optical trap in air. We work in a dual-beam horizontal configuration with tunable laser intensity \cite{Reich2023}, allowing us to systematically study the laser's effect on the charging dynamics. Working with high-purity ($>99.9 \%$) amorphous silica, we find charging dynamics that are indicative of a two-photon process. Based thereon, we develop a mathematical model that closely reproduces our experimental charging \textit{vs.}~time curves. With straightforward modifications, the use of optical tweezers to simultaneously charge and measure the charge of individual microscopic objects should be applicable to a wide variety of situations, ranging from fundamental studies of electronic properties to charge dynamics in aerosol physics and cloud electrification. 

The experimental setup is sketched in Fig.~\ref{fig:experimental_setup}. We use a continuous wave 532 nm laser (Laser Quantum, opus 532, 3W) to trap a sub-micron, amorphous silica (SiO\textsubscript{2}) sphere (Cospheric SiO2MS-2.0, nominal radius $r\approx0.35$ \textmu{m}, density 2.0 g/cm\textsuperscript{3}).  
The initially free-space beam passes through a polarizing beam splitter (PBS) for easy intensity tuning and is then divided by a second PBS into two arms of equal power. Subsequently, each arm is coupled into a single-mode optical fiber (Thorlabs, FPC021) to facilitate beam transport, polarization control, and spatial filtering. The two beams are brought back to free space and expanded to $\sim$0.5~cm by two collimators (Thorlabs, TC25APC-532) shortly before they enter the grounded aluminum experimental chamber held at atmospheric pressure. Two lenses (Thorlabs, A240TM-A, focal lengths 8 mm) inside the chamber focus the orthogonally polarized beams to form a counter-propagating optical trap (numerical apertures 0.29, $\sim$6 \textmu{m} separation between foci). The lens holders are machined from copper and simultaneously serve as electrodes to apply an electric field along the optical axis. The voltage applied is typically 250 V at frequency $f_{dr}$ = 2 kHz, yielding an AC electric field at the particle of $E_0\approx$ $8.5\times10^3$ V/m. 

If the particle is charged, the electric field ``shakes'' it on its ``optical spring'' at the driving frequency. Outside the chamber, an inline beam splitter picks off a fraction of the light scattered from the shaking particle and diverts it to a photodiode (Thorlabs, PDAPC2), from which we record continuously at 1 MHz. The measured fluctuations are proportional to the particle's position along the beam axis. We convert this position signal into a real-time charge measurement following the method of Ref.~\cite{Ricci2019}. In brief, this entails comparing the electrically driven contribution to the power spectral density (PSD) of the particle's motion at the driving frequency ($S_{el}$) to what the PSD would be at the same frequency due to thermal motion alone ($S_{th}$). The mathematical expression for the particle charge, $Q$, is

\begin{equation}
    Q = \sqrt\frac{8 k_B T \gamma S_{el}(w_{dr})} {E_0^2 \tau S_{th}(w_{dr})},
    \label{eq:charge equation}
\end{equation}
where $k_B$, $T$ and $\gamma$ are the Boltzmann constant, the temperature and the drag coefficient, respectively, and $\tau$ is the time window over which the PSD is calculated. The full derivation of Eq.~\ref{eq:charge equation} can be found in Ref.~\cite{Ricci2019}; for the reader's convenience we provide a sketch in the Supplemental Material \cite{SupplMat}. To extract the sign of the particle's charge, we additionally monitor the phase of the position oscillations relative to the applied electric field. For computational efficiency, we do not compute the full PSD in every time window but instead use a computational lock-in approach where we measure the charge peak and estimate $S_{th}$, ignoring other frequency components. $S_{el}$ is measured by computing the squared Fourier component of the raw signal at the driving frequency and subtracting $S_{th}$, while $S_{th}$ is estimated with the geometric average of the components of the signal $\pm$300 Hz on either side (far enough from the charge peak so that $S_{th}$ does not change with $Q$), as illustrated in Fig.~\ref{fig:measurement_principle}(a). For these experiments, we used $\tau = 0.2$ s, resulting in 5 charge measurements per second. To ensure zero particle charge at the beginning of a measurement, we use a soft X-ray source (Hamamatsu L9491) to ionize the air around the particle (without irradiating the particle itself), which enhances the local air conductivity and neutralizes the particle within $\sim$$10^{-2}$ s \cite{Hinds1999}. Fig.~\ref{fig:measurement_principle}(b) illustrates the decaying oscillations when the x-ray source is switched on, leaving only the particle's Brownian motion in the trap after the discharge is complete. For small $Q$ and low laser intensities, we resolve the particle's electric charge in elementary units, showing every charging (or discharging) event as a discrete step of one elementary charge, $|e|$ (as illustrated for a short duration in Fig.~\ref{fig:measurement_principle}(c)).

\begin{figure}[ht!]
\centering
\includegraphics{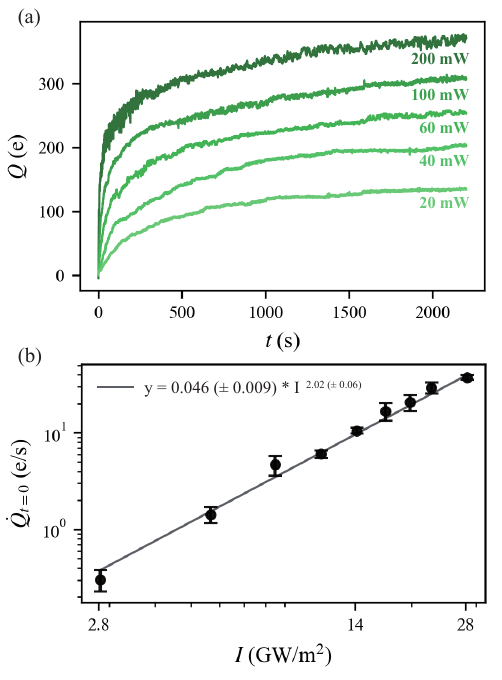}
\caption{ \label{fig:charge_evolution} Charge evolution. (a) First 2200 seconds of the charging curves of the particle at different laser powers applied to the optical trap. (b) Initial charging rate (up to $\sim$150 e) versus laser intensity $I$. The square-law dependence shows that the particle is ionized by a two-photon process.
}
\end{figure}

Fig.~\ref{fig:charge_evolution}(a) shows the main experimental results, which are charge \textit{vs.}~time curves for different powers of the trapping laser. As can be seen, the charging evolution depends on the laser power applied to the optical trap as well as on time. At first, the charging rate is quite linear, but as time progresses, it slows down considerably. In Fig.~\ref{fig:charge_evolution}(b), we plot the initial charging rate as a function of the laser intensity, $I$, with error bars calculated in each case from three independent runs. This data fits precisely to $I^2$---a clear signature of some kind of two-photon process associated with the charging mechanism. Although the charging rate of each curve slows down over time, data like those shown in Fig.~\ref{fig:charge_evolution} do not fit to typical saturating functions---\textit{e.g.}, a saturating exponential, hyperbolic tangent, or rational function.  At very long times, \textit{i.e.}~much longer than the 2200 s shown in Fig.~\ref{fig:charge_evolution}, we do encounter an upper limit, but this is due to the fact that the particle becomes so charged that it causes breakdown in the surrounding air---see the Supplemental Material \cite{SupplMat}.  

While it is evident from Fig.~\ref{fig:charge_evolution}(b) that two-photon absorption is involved in the observed charging, it is not immediately clear how. The valence band maximum of SiO$_2$ is typically reported to be in the range of $\sim$10 eV below the vacuum level \cite{Ishii1990, Fujimura2016, OBrien2000, Fulton2006}; hence, two of our green photons (2$h\nu\approx4.66$ eV) are insufficient for direct liberation to air from the valence band. On the other hand, there can be in-gap  states in amorphous SiO$_2$ above the valence band maximum. Assuming the liberated electrons come from these states, we can imagine two physical interpretations that explain our observations. As it turns out, both result in the same logarithmic fitting equation, which describes our data well. For brevity's sake, we sketch only the essential physics behind these models here; full derivations are in the Supplemental Material \cite{SupplMat}. 

In the first interpretation, we suppose that some relevant initial electronic state in the bandgap is at an energy below free, $\epsilon$, deeper than $2h\nu$, hence two-photons are insufficient for direct liberation. Even so, if the deficit ($\delta=\epsilon-2h\nu$) is not too large, it is plausible that two-photon excitation can occur to some intermediate state (\textit{e.g.}, the conduction band), which could then be followed by full liberation via thermal emission. Initially, the barrier to thermal emission would be the energy deficit, $\delta$. However, as the charge of the particle grows, the additional electrostatic barrier, $\tfrac{Qe}{4\pi\epsilon_0 r}$, must be overcome. In the case that the fraction of thermally escaping electrons is small compared to the steady-state population in the intermediate state, it can be shown that this scenario leads to the charging rate equation,
\begin{equation}
\frac{dQ}{dt} \propto I^2 e^{-b_t\tfrac{Qe}{4\pi\epsilon_0 rk_BT}},
\label{eq:thermal}
\end{equation}
where $k_B$ is Boltzmann's constant, $T$ is temperature, and $b_t$ should be equal to unity (more on this later).  

In the second scenario, we imagine that there is a \textit{distribution} of occupied electronic states in the bandgap as a function of the energetic depth, $\epsilon$, and that electrons of sufficiently shallow energies in this distribution can be directly liberated via the two-photon process. We define the cumulative distribution of these occupied states, \textit{i.e.}, the total number between 0 and $\epsilon$, as $N(\epsilon)$. If the particle has charge $Q$, then the total number of electrons that can possibly be liberated is $N\big{(}2h\nu-\tfrac{Qe}{4\pi\epsilon_0 r}\big{)}$. The inclusion of the Coulombic term now indicates that the two-photon energy must be sufficient to liberate from the material and overcome the electrostatic well of the charged particle. In addition to the $I^2$ dependence, the charging rate should be proportional to this number. Hence, in this scenario the rate equation is 

\begin{equation}
\frac{dQ}{dt} \propto I^2N\big{(}2h\nu-\tfrac{Qe}{4\pi\epsilon_0 r} \big{)}.
\label{eq:distribution}
\end{equation}
As can be seen, if the cumulative distribution of inter-gap states is exponential, \textit{i.e.}, if $N(\epsilon)\propto e^{b_d\epsilon}$ (with $b_d$ an unknown coefficient not necessarily related to $k_BT$) then Eq.~\ref{eq:thermal} and Eq.~\ref{eq:distribution} are functionally identical. 

\begin{figure}[ht!]
\centering
\includegraphics{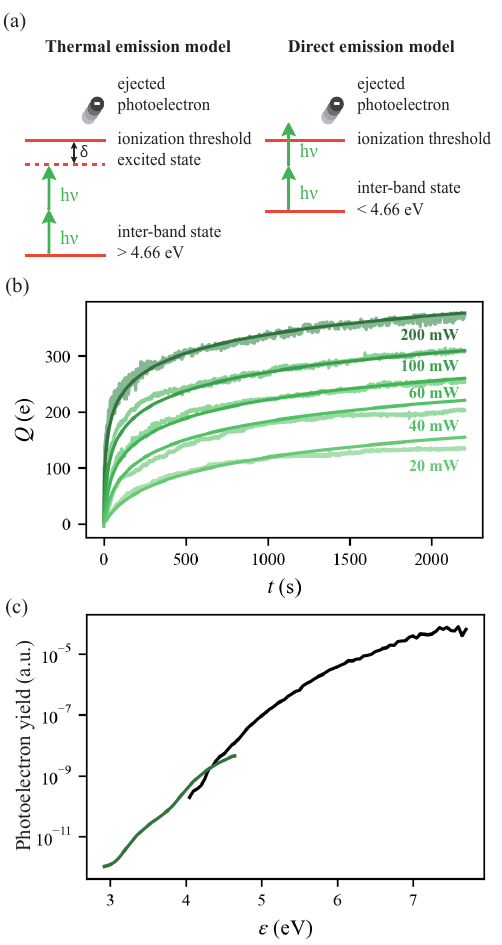}
\caption{ \label{fig:two models} Illustration of the two proposed models, experimental and modeled charging curves and photoelectron yield spectrum of our sample. (a) In the thermal emission model (left), an electron is excited from a deeper ($>$ 4.66 eV) trap state by the simultaneous absorption of two green photons. From this excited state, the remaining ionization barrier $\delta$ is overcome thermally, resulting in the emission of an electron. In the direct emission model (right) the energy of two photons is enough to directly eject an electron from a trap state shallower than 4.66 eV. (b) Experimental data (background) and model charging curves predicted from Eq. \ref{eq:model equation} (foreground). (c) Photoelectron yield spectroscopy data for our sample in bulk powder form (black). The green curve shows the
(rescaled) charging rate versus ionization energy for one of our 200 mW curves from the optical tweezers experiment. The data show the same exponential growth as the PYS data and reveals ionization events at energies as low as 3 eV.
}
\end{figure}

In either case, the differential equation is easily solved, subject to the initial condition $Q(0)=0$ as enforced by our initial discharge of the sphere.  The solution is of the form

\begin{equation}
    Q(t) = \frac{4 \pi \epsilon_0 r}{B e} \ln \bigg{(}\frac{e}{4 \pi \epsilon_0 r}A B I^2 t + 1 \bigg{)}.
\label{eq:model equation}
\end{equation}
Here, $A$ and $B$ are fitting constants related to the coefficient in front of $I^2$ and the proportionality factors in the exponential ($b_t$ or $b_d$, dependent on the model). As we explain in the Supplemental Material, care has to be taken regarding the exact definitions of $A$ and $B$ in each case \cite{SupplMat}. As we show in Fig.~\ref{fig:two models}(b), this functional form fits our data well, recovering all of the essential features. In particular, we observe that the curves are proportional to $I^2$ initially, and that the charging rate slows down but does not show obvious saturation behavior.

Which physical interpretation is correct? The thermal emission model is appealing in that it offers a first-principles motivation for the exponential in the rate equation (\textit{i.e.}, the Boltzmann factor), yet here is where it also has a significant drawback. As we show in the Supplemental Material \cite{SupplMat}, the $b_t$ parameter extracted for the thermal model is different from unity by about a factor of $\times10$. This could in principle be consistent with an elevated temperature of the particle; however, due to the low absorption coefficient of silica at the trapping wavelength, it should remain very close to ambient temperature $\sim 23 \text{ } ^\circ\text{C}$ (see Supplemental Material \cite{SupplMat}). On the face of it, the direct-emission model would seem to have the disadvantage that it requires an exponentially distributed population of states in the gap without a first principles motivation.  Perhaps ironically, this is where the direct-emission becomes more tenable. In Fig.~\ref{fig:two models}(c), we present photoelectron yield spectroscopy (PYS) data for our samples in bulk powder form.  These data were taken on a dedicated PYS-instrument that is independent from our main optical tweezers experiment (see Supplemental Materials for details on PYS measurements \cite{SupplMat}). In this data, tunable UV light illuminates the sample and the resulting photocurrent (yield) is measured. The data show that the yield grows exponentially with the single-photon energy, which  is consistent with the assumption that the cumulative distribution of states, $N(\epsilon)$, grows exponentially with $\epsilon$. Moreover, while the PYS data is only reliable down to about 4 eV, we can extract an effectively identical yield \textit{vs.}~energy curve from our single sphere data in our optical trap (\textit{i.e.}, from the curves in Fig.~\ref{fig:charge_evolution}(a)), which reveals it has the same factor in the exponential ($b_d$) and extends the range down to as little as 3 eV (green curve Fig.~\ref{fig:two models}(c); see Supplemental Information for details \cite{SupplMat}).

Photoemission curves such as those shown in Fig.~\ref{fig:two models}(c) with broad exponential dependencies in the band gap are routinely observed in atomically disordered materials, and are commonly attributed to the non-equilibrium electronic landscape due to structural disorder, \textit{i.e.}, not due to thermal equilibrium \cite{Wager2017, Singh2003}. The existence of exponential tail states for SiO$_2$ is also suggested from photoemission measurement in the Supplemental Information \cite{SupplMat}. Hence, there is precedent to view our data with this picture in mind. Practically, however, we are not able to distinguish between the two physical interpretations using our current optical tweezers setup. We cannot exclude the possibility that the thermal picture is ultimately correct, but some additional ingredient we aren't considering makes the factor $b_t$ different from unity. For instance, the distribution of charges could deviate from spherical due to the presence of the AC electric field, thereby facilitating electron escape. As the thermal emission scenario is sensitive to temperature in a clearly defined way, whereas the direct emission scenario is not, a future direction for experiments could be to incorporate temperature control, \textit{e.g.}, via an additional deep IR laser that, in contrast to our green laser, is significantly absorbed by and can therefore heat the particle. Beyond either of the proposed mechanisms, additional contributions, \textit{e.g.}, lowering of the escape barrier via the Poole–Frenkel effect \cite{Frenkel1938} caused by the applied field, might also be considered. Yet, the present functional form already fits the data quite well.

The simultaneous capacity to levitate a micron-scale (or smaller) particle, measure its charge with electron-scale resolution, and charge it to the maximum amount it can sustain before dielectric breakdown occurs is significant, offering potential uses across various fields. Our data already show that it offers a way to probe electronic states, creating a retarding potential that is relatively easy to understand and account for. Of special interest to the authors is the potential to study charging dynamics with particles relevant to aerosol and cloud electrification. Understanding how these particles acquire and---perhaps as importantly---\textit{lose} charge (\textit{e.g.}, as demonstrated in the Supplemental Material \cite{SupplMat}) could provide new insights into some of the most persistent questions in these topics, for instance lightning initiation. Furthermore, it is easy to envision straightforward additions to the technique that expand its usefulness. Existing experiments with optical tweezers already show that lower energy photons (\textit{e.g.}, in the near IR \cite{Ricci2019, Beunis2010, Beunis2012}) generally avoid the charging mechanism we observe; hence by overlapping one ``trapping'' laser (near IR) and one ``charging'' laser (such as ours), a particle's charge could be brought to a desired value and then ``set free'' to evolve in different environmental conditions. Moreover, with a sufficiently high energy UV laser, the charging mechanism need not be from a two-photon process, and as we suggested earlier an overlapping deep IR laser could be used to heat the particle to enable temperature-dependent studies. These possibilities position optically levitated particles as a unique platform for probing complex, charge-related behavior of microscopic particles across a wide range of disciplines, with a level of control, specificity, and resolution that is difficult to achieve by other means.

\vspace{1em}
We thank Todor Asenov and Abdulhamid Baghdadi for their outstanding technical support and Dr. Michael Gleichweit and Mercede Azizbaig Mohajer for the helpful discussions. This project has received funding from the European Research Council (ERC) under the European Union’s Horizon 2020 research and innovation programme (Grant agreements No.~949120 and No.~805041) and the Swiss National Science Foundation (SNSF, project 200021-236446). This research was supported by the Scientific Service Units of The Institute of Science and Technology Austria (ISTA) through resources provided by the Miba Machine Shop and the Scientific Computing service unit.

\bibliographystyle{apsrev4-2}

\begin{thebibliography}{26}%
\makeatletter
\providecommand \@ifxundefined [1]{%
 \@ifx{#1\undefined}
}%
\providecommand \@ifnum [1]{%
 \ifnum #1\expandafter \@firstoftwo
 \else \expandafter \@secondoftwo
 \fi
}%
\providecommand \@ifx [1]{%
 \ifx #1\expandafter \@firstoftwo
 \else \expandafter \@secondoftwo
 \fi
}%
\providecommand \natexlab [1]{#1}%
\providecommand \enquote  [1]{``#1''}%
\providecommand \bibnamefont  [1]{#1}%
\providecommand \bibfnamefont [1]{#1}%
\providecommand \citenamefont [1]{#1}%
\providecommand \href@noop [0]{\@secondoftwo}%
\providecommand \href [0]{\begingroup \@sanitize@url \@href}%
\providecommand \@href[1]{\@@startlink{#1}\@@href}%
\providecommand \@@href[1]{\endgroup#1\@@endlink}%
\providecommand \@sanitize@url [0]{\catcode `\\12\catcode `\$12\catcode `\&12\catcode `\#12\catcode `\^12\catcode `\_12\catcode `\%12\relax}%
\providecommand \@@startlink[1]{}%
\providecommand \@@endlink[0]{}%
\providecommand \url  [0]{\begingroup\@sanitize@url \@url }%
\providecommand \@url [1]{\endgroup\@href {#1}{\urlprefix }}%
\providecommand \urlprefix  [0]{URL }%
\providecommand \Eprint [0]{\href }%
\providecommand \doibase [0]{https://doi.org/}%
\providecommand \selectlanguage [0]{\@gobble}%
\providecommand \bibinfo  [0]{\@secondoftwo}%
\providecommand \bibfield  [0]{\@secondoftwo}%
\providecommand \translation [1]{[#1]}%
\providecommand \BibitemOpen [0]{}%
\providecommand \bibitemStop [0]{}%
\providecommand \bibitemNoStop [0]{.\EOS\space}%
\providecommand \EOS [0]{\spacefactor3000\relax}%
\providecommand \BibitemShut  [1]{\csname bibitem#1\endcsname}%
\let\auto@bib@innerbib\@empty
\bibitem [{\citenamefont {Merrill}\ \emph {et~al.}(2009)\citenamefont {Merrill}, \citenamefont {Sainis},\ and\ \citenamefont {Dufresne}}]{Merrill2009}%
  \BibitemOpen
  \bibfield  {author} {\bibinfo {author} {\bibfnamefont {J.~W.}\ \bibnamefont {Merrill}}, \bibinfo {author} {\bibfnamefont {S.~K.}\ \bibnamefont {Sainis}},\ and\ \bibinfo {author} {\bibfnamefont {E.~R.}\ \bibnamefont {Dufresne}},\ }\href@noop {} {\bibfield  {journal} {\bibinfo  {journal} {Phys. Rev. Lett.}\ }\textbf {\bibinfo {volume} {103}},\ \bibinfo {pages} {138301} (\bibinfo {year} {2009})}\BibitemShut {NoStop}%
\bibitem [{\citenamefont {Mitchem}\ and\ \citenamefont {Reid}(2008)}]{Mitchem2008}%
  \BibitemOpen
  \bibfield  {author} {\bibinfo {author} {\bibfnamefont {L.}~\bibnamefont {Mitchem}}\ and\ \bibinfo {author} {\bibfnamefont {J.~P.}\ \bibnamefont {Reid}},\ }\href@noop {} {\bibfield  {journal} {\bibinfo  {journal} {Chem. Soc. Rev.}\ }\textbf {\bibinfo {volume} {37}},\ \bibinfo {pages} {756} (\bibinfo {year} {2008})}\BibitemShut {NoStop}%
\bibitem [{\citenamefont {Mohajer}\ \emph {et~al.}(2025)\citenamefont {Mohajer}, \citenamefont {Basuri}, \citenamefont {Evdokimov}, \citenamefont {David}, \citenamefont {Zindel}, \citenamefont {Miliordos},\ and\ \citenamefont {Signorell}}]{Mohajer2025}%
  \BibitemOpen
  \bibfield  {author} {\bibinfo {author} {\bibfnamefont {M.~A.}\ \bibnamefont {Mohajer}}, \bibinfo {author} {\bibfnamefont {P.}~\bibnamefont {Basuri}}, \bibinfo {author} {\bibfnamefont {A.}~\bibnamefont {Evdokimov}}, \bibinfo {author} {\bibfnamefont {G.}~\bibnamefont {David}}, \bibinfo {author} {\bibfnamefont {D.}~\bibnamefont {Zindel}}, \bibinfo {author} {\bibfnamefont {E.}~\bibnamefont {Miliordos}},\ and\ \bibinfo {author} {\bibfnamefont {R.}~\bibnamefont {Signorell}},\ }\href {https://doi.org/10.1126/science.adv2362} {\bibfield  {journal} {\bibinfo  {journal} {Science}\ }\textbf {\bibinfo {volume} {388}},\ \bibinfo {pages} {1426} (\bibinfo {year} {2025})}\BibitemShut {NoStop}%
\bibitem [{\citenamefont {Sullivan}\ \emph {et~al.}(2020)\citenamefont {Sullivan}, \citenamefont {Boyer-Chelmo}, \citenamefont {Gorkowski},\ and\ \citenamefont {Beydoun}}]{Sullivan2020}%
  \BibitemOpen
  \bibfield  {author} {\bibinfo {author} {\bibfnamefont {R.~C.}\ \bibnamefont {Sullivan}}, \bibinfo {author} {\bibfnamefont {H.}~\bibnamefont {Boyer-Chelmo}}, \bibinfo {author} {\bibfnamefont {K.}~\bibnamefont {Gorkowski}},\ and\ \bibinfo {author} {\bibfnamefont {H.}~\bibnamefont {Beydoun}},\ }\href@noop {} {\bibfield  {journal} {\bibinfo  {journal} {Acc. Chem. Res.}\ }\textbf {\bibinfo {volume} {53}},\ \bibinfo {pages} {2498} (\bibinfo {year} {2020})}\BibitemShut {NoStop}%
\bibitem [{\citenamefont {Ruffner}\ and\ \citenamefont {Grier}(2012)}]{Ruffner2012}%
  \BibitemOpen
  \bibfield  {author} {\bibinfo {author} {\bibfnamefont {D.~B.}\ \bibnamefont {Ruffner}}\ and\ \bibinfo {author} {\bibfnamefont {D.~G.}\ \bibnamefont {Grier}},\ }\href@noop {} {\bibfield  {journal} {\bibinfo  {journal} {Phys. Rev. Lett.}\ }\textbf {\bibinfo {volume} {108}},\ \bibinfo {pages} {173602} (\bibinfo {year} {2012})}\BibitemShut {NoStop}%
\bibitem [{\citenamefont {Rubinsztein-Dunlop}\ \emph {et~al.}(2015)\citenamefont {Rubinsztein-Dunlop}, \citenamefont {Stilgoe}, \citenamefont {Preece}, \citenamefont {Bui},\ and\ \citenamefont {Nieminen}}]{Rubinsztein-Dunlop2015}%
  \BibitemOpen
  \bibfield  {author} {\bibinfo {author} {\bibfnamefont {H.}~\bibnamefont {Rubinsztein-Dunlop}}, \bibinfo {author} {\bibfnamefont {A.~B.}\ \bibnamefont {Stilgoe}}, \bibinfo {author} {\bibfnamefont {D.}~\bibnamefont {Preece}}, \bibinfo {author} {\bibfnamefont {A.}~\bibnamefont {Bui}},\ and\ \bibinfo {author} {\bibfnamefont {T.~A.}\ \bibnamefont {Nieminen}},\ }in\ \href@noop {} {\emph {\bibinfo {booktitle} {Photonics}}}\ (\bibinfo  {publisher} {"John Wiley \& Sons, Inc."},\ \bibinfo {address} {Hoboken, NJ, USA},\ \bibinfo {year} {2015})\ pp.\ \bibinfo {pages} {287--339}\BibitemShut {NoStop}%
\bibitem [{\citenamefont {Millikan.}(1913)}]{Millikan1913}%
  \BibitemOpen
  \bibfield  {author} {\bibinfo {author} {\bibfnamefont {R.~A.}\ \bibnamefont {Millikan.}},\ }\href@noop {} {\bibfield  {journal} {\bibinfo  {journal} {Phys. Rev.}\ }\textbf {\bibinfo {volume} {2}},\ \bibinfo {pages} {109} (\bibinfo {year} {1913})}\BibitemShut {NoStop}%
\bibitem [{\citenamefont {Beunis}\ \emph {et~al.}(2010)\citenamefont {Beunis}, \citenamefont {Strubbe}, \citenamefont {Verboven}, \citenamefont {Neyts},\ and\ \citenamefont {Petrov}}]{Beunis2010}%
  \BibitemOpen
  \bibfield  {author} {\bibinfo {author} {\bibfnamefont {F.}~\bibnamefont {Beunis}}, \bibinfo {author} {\bibfnamefont {F.}~\bibnamefont {Strubbe}}, \bibinfo {author} {\bibfnamefont {B.}~\bibnamefont {Verboven}}, \bibinfo {author} {\bibfnamefont {K.}~\bibnamefont {Neyts}},\ and\ \bibinfo {author} {\bibfnamefont {D.}~\bibnamefont {Petrov}},\ }in\ \href@noop {} {\emph {\bibinfo {booktitle} {Complex Light and Optical Forces {IV}}}},\ \bibinfo {editor} {edited by\ \bibinfo {editor} {\bibfnamefont {E.~J.}\ \bibnamefont {Galvez}}, \bibinfo {editor} {\bibfnamefont {D.~L.}\ \bibnamefont {Andrews}},\ and\ \bibinfo {editor} {\bibfnamefont {J.}~\bibnamefont {Gl{\"u}ckstad}}}\ (\bibinfo  {publisher} {SPIE},\ \bibinfo {year} {2010})\BibitemShut {NoStop}%
\bibitem [{\citenamefont {Beunis}\ \emph {et~al.}(2012)\citenamefont {Beunis}, \citenamefont {Strubbe}, \citenamefont {Neyts},\ and\ \citenamefont {Petrov}}]{Beunis2012}%
  \BibitemOpen
  \bibfield  {author} {\bibinfo {author} {\bibfnamefont {F.}~\bibnamefont {Beunis}}, \bibinfo {author} {\bibfnamefont {F.}~\bibnamefont {Strubbe}}, \bibinfo {author} {\bibfnamefont {K.}~\bibnamefont {Neyts}},\ and\ \bibinfo {author} {\bibfnamefont {D.}~\bibnamefont {Petrov}},\ }\bibfield  {journal} {\bibinfo  {journal} {Physical Review Letters}\ }\textbf {\bibinfo {volume} {108}},\ \href {https://doi.org/10.1103/PhysRevLett.108.016101} {10.1103/PhysRevLett.108.016101} (\bibinfo {year} {2012})\BibitemShut {NoStop}%
\bibitem [{\citenamefont {Schreuer}\ \emph {et~al.}(2018{\natexlab{a}})\citenamefont {Schreuer}, \citenamefont {Vandewiele}, \citenamefont {Strubbe}, \citenamefont {Neyts},\ and\ \citenamefont {Beunis}}]{Schreuer2018}%
  \BibitemOpen
  \bibfield  {author} {\bibinfo {author} {\bibfnamefont {C.}~\bibnamefont {Schreuer}}, \bibinfo {author} {\bibfnamefont {S.}~\bibnamefont {Vandewiele}}, \bibinfo {author} {\bibfnamefont {F.}~\bibnamefont {Strubbe}}, \bibinfo {author} {\bibfnamefont {K.}~\bibnamefont {Neyts}},\ and\ \bibinfo {author} {\bibfnamefont {F.}~\bibnamefont {Beunis}},\ }\href@noop {} {\bibfield  {journal} {\bibinfo  {journal} {J. Colloid Interface Sci.}\ }\textbf {\bibinfo {volume} {515}},\ \bibinfo {pages} {248} (\bibinfo {year} {2018}{\natexlab{a}})}\BibitemShut {NoStop}%
\bibitem [{\citenamefont {Schreuer}\ \emph {et~al.}(2018{\natexlab{b}})\citenamefont {Schreuer}, \citenamefont {Vandewiele}, \citenamefont {Brans}, \citenamefont {Strubbe}, \citenamefont {Neyts},\ and\ \citenamefont {Beunis}}]{Schreuer2018-2}%
  \BibitemOpen
  \bibfield  {author} {\bibinfo {author} {\bibfnamefont {C.}~\bibnamefont {Schreuer}}, \bibinfo {author} {\bibfnamefont {S.}~\bibnamefont {Vandewiele}}, \bibinfo {author} {\bibfnamefont {T.}~\bibnamefont {Brans}}, \bibinfo {author} {\bibfnamefont {F.}~\bibnamefont {Strubbe}}, \bibinfo {author} {\bibfnamefont {K.}~\bibnamefont {Neyts}},\ and\ \bibinfo {author} {\bibfnamefont {F.}~\bibnamefont {Beunis}},\ }\href@noop {} {\bibfield  {journal} {\bibinfo  {journal} {J. Appl. Phys.}\ }\textbf {\bibinfo {volume} {123}},\ \bibinfo {pages} {015105} (\bibinfo {year} {2018}{\natexlab{b}})}\BibitemShut {NoStop}%
\bibitem [{\citenamefont {Ricci}\ \emph {et~al.}(2019)\citenamefont {Ricci}, \citenamefont {Cuairan}, \citenamefont {Conangla}, \citenamefont {Schell},\ and\ \citenamefont {Quidant}}]{Ricci2019}%
  \BibitemOpen
  \bibfield  {author} {\bibinfo {author} {\bibfnamefont {F.}~\bibnamefont {Ricci}}, \bibinfo {author} {\bibfnamefont {M.~T.}\ \bibnamefont {Cuairan}}, \bibinfo {author} {\bibfnamefont {G.~P.}\ \bibnamefont {Conangla}}, \bibinfo {author} {\bibfnamefont {A.~W.}\ \bibnamefont {Schell}},\ and\ \bibinfo {author} {\bibfnamefont {R.}~\bibnamefont {Quidant}},\ }\href {https://doi.org/10.1021/acs.nanolett.9b00082} {\bibfield  {journal} {\bibinfo  {journal} {Nano Letters}\ }\textbf {\bibinfo {volume} {19}},\ \bibinfo {pages} {6711} (\bibinfo {year} {2019})}\BibitemShut {NoStop}%
\bibitem [{\citenamefont {Zhu}\ \emph {et~al.}(2023)\citenamefont {Zhu}, \citenamefont {Fu}, \citenamefont {Gao}, \citenamefont {Li}, \citenamefont {Chen}, \citenamefont {Wang}, \citenamefont {Chen},\ and\ \citenamefont {Hu}}]{Zhu2023}%
  \BibitemOpen
  \bibfield  {author} {\bibinfo {author} {\bibfnamefont {S.}~\bibnamefont {Zhu}}, \bibinfo {author} {\bibfnamefont {Z.}~\bibnamefont {Fu}}, \bibinfo {author} {\bibfnamefont {X.}~\bibnamefont {Gao}}, \bibinfo {author} {\bibfnamefont {C.}~\bibnamefont {Li}}, \bibinfo {author} {\bibfnamefont {Z.}~\bibnamefont {Chen}}, \bibinfo {author} {\bibfnamefont {Y.}~\bibnamefont {Wang}}, \bibinfo {author} {\bibfnamefont {X.}~\bibnamefont {Chen}},\ and\ \bibinfo {author} {\bibfnamefont {H.}~\bibnamefont {Hu}},\ }\href {https://doi.org/10.1364/PRJ.475793} {\bibfield  {journal} {\bibinfo  {journal} {Photon. Res.}\ }\textbf {\bibinfo {volume} {11}},\ \bibinfo {pages} {279} (\bibinfo {year} {2023})}\BibitemShut {NoStop}%
\bibitem [{\citenamefont {Pesce}\ \emph {et~al.}(2015)\citenamefont {Pesce}, \citenamefont {Rusciano}, \citenamefont {Zito},\ and\ \citenamefont {Sasso}}]{Pesce2015}%
  \BibitemOpen
  \bibfield  {author} {\bibinfo {author} {\bibfnamefont {G.}~\bibnamefont {Pesce}}, \bibinfo {author} {\bibfnamefont {G.}~\bibnamefont {Rusciano}}, \bibinfo {author} {\bibfnamefont {G.}~\bibnamefont {Zito}},\ and\ \bibinfo {author} {\bibfnamefont {A.}~\bibnamefont {Sasso}},\ }\href@noop {} {\bibfield  {journal} {\bibinfo  {journal} {Opt. Express}\ }\textbf {\bibinfo {volume} {23}},\ \bibinfo {pages} {9363} (\bibinfo {year} {2015})}\BibitemShut {NoStop}%
\bibitem [{\citenamefont {Marmolejo}\ \emph {et~al.}(2021)\citenamefont {Marmolejo}, \citenamefont {Urquiza-González}, \citenamefont {Isaksson}, \citenamefont {Johansson}, \citenamefont {Méndez-Fragoso},\ and\ \citenamefont {Hanstorp}}]{Marmolejo2021}%
  \BibitemOpen
  \bibfield  {author} {\bibinfo {author} {\bibfnamefont {J.~T.}\ \bibnamefont {Marmolejo}}, \bibinfo {author} {\bibfnamefont {M.}~\bibnamefont {Urquiza-González}}, \bibinfo {author} {\bibfnamefont {O.}~\bibnamefont {Isaksson}}, \bibinfo {author} {\bibfnamefont {A.}~\bibnamefont {Johansson}}, \bibinfo {author} {\bibfnamefont {R.}~\bibnamefont {Méndez-Fragoso}},\ and\ \bibinfo {author} {\bibfnamefont {D.}~\bibnamefont {Hanstorp}},\ }\bibfield  {journal} {\bibinfo  {journal} {Scientific Reports}\ }\textbf {\bibinfo {volume} {11}},\ \href {https://doi.org/10.1038/s41598-021-89714-2} {10.1038/s41598-021-89714-2} (\bibinfo {year} {2021})\BibitemShut {NoStop}%
\bibitem [{\citenamefont {Ashkin}\ and\ \citenamefont {Dziedzic}(1976)}]{Ashkin1976}%
  \BibitemOpen
  \bibfield  {author} {\bibinfo {author} {\bibfnamefont {A.}~\bibnamefont {Ashkin}}\ and\ \bibinfo {author} {\bibfnamefont {J.~M.}\ \bibnamefont {Dziedzic}},\ }\href@noop {} {\bibfield  {journal} {\bibinfo  {journal} {Phys. Rev. Lett.}\ }\textbf {\bibinfo {volume} {36}},\ \bibinfo {pages} {267} (\bibinfo {year} {1976})}\BibitemShut {NoStop}%
\bibitem [{\citenamefont {Reich}\ \emph {et~al.}(2023)\citenamefont {Reich}, \citenamefont {Gleichweit}, \citenamefont {David}, \citenamefont {Leemann},\ and\ \citenamefont {Signorell}}]{Reich2023}%
  \BibitemOpen
  \bibfield  {author} {\bibinfo {author} {\bibfnamefont {O.}~\bibnamefont {Reich}}, \bibinfo {author} {\bibfnamefont {M.~J.}\ \bibnamefont {Gleichweit}}, \bibinfo {author} {\bibfnamefont {G.}~\bibnamefont {David}}, \bibinfo {author} {\bibfnamefont {N.}~\bibnamefont {Leemann}},\ and\ \bibinfo {author} {\bibfnamefont {R.}~\bibnamefont {Signorell}},\ }\href@noop {} {\bibfield  {journal} {\bibinfo  {journal} {Environ. Sci. Atmos.}\ }\textbf {\bibinfo {volume} {3}},\ \bibinfo {pages} {695} (\bibinfo {year} {2023})}\BibitemShut {NoStop}%
\bibitem [{Sup()}]{SupplMat}%
  \BibitemOpen
  \href@noop {} {\emph {\bibinfo {title} {{See Supplemental Material at [URL will be inserted by publisher].}}}}\BibitemShut {Stop}%
\bibitem [{\citenamefont {Hinds}(1999)}]{Hinds1999}%
  \BibitemOpen
  \bibfield  {author} {\bibinfo {author} {\bibfnamefont {W.~C.}\ \bibnamefont {Hinds}},\ }\href {https://ebookcentral-proquest-com.uaccess.univie.ac.at/lib/univie/detail.action?docID=1120423} {\emph {\bibinfo {title} {Aerosol technology: properties, behavior, and measurement of airborne particles}}}\ (\bibinfo  {publisher} {John Wiley \& Sons, Incorporated},\ \bibinfo {year} {1999})\BibitemShut {NoStop}%
\bibitem [{\citenamefont {Ishii}\ \emph {et~al.}(1990)\citenamefont {Ishii}, \citenamefont {Masuda},\ and\ \citenamefont {Harada}}]{Ishii1990}%
  \BibitemOpen
  \bibfield  {author} {\bibinfo {author} {\bibfnamefont {H.}~\bibnamefont {Ishii}}, \bibinfo {author} {\bibfnamefont {S.}~\bibnamefont {Masuda}},\ and\ \bibinfo {author} {\bibfnamefont {Y.}~\bibnamefont {Harada}},\ }\href {https://doi.org/https://doi.org/10.1016/0039-6028(90)90224-V} {\bibfield  {journal} {\bibinfo  {journal} {Surface Science}\ }\textbf {\bibinfo {volume} {239}},\ \bibinfo {pages} {222} (\bibinfo {year} {1990})}\BibitemShut {NoStop}%
\bibitem [{\citenamefont {Fujimura}\ \emph {et~al.}(2016)\citenamefont {Fujimura}, \citenamefont {Ohta}, \citenamefont {Makihara},\ and\ \citenamefont {Miyazaki}}]{Fujimura2016}%
  \BibitemOpen
  \bibfield  {author} {\bibinfo {author} {\bibfnamefont {N.}~\bibnamefont {Fujimura}}, \bibinfo {author} {\bibfnamefont {A.}~\bibnamefont {Ohta}}, \bibinfo {author} {\bibfnamefont {K.}~\bibnamefont {Makihara}},\ and\ \bibinfo {author} {\bibfnamefont {S.}~\bibnamefont {Miyazaki}},\ }\href@noop {} {\bibfield  {journal} {\bibinfo  {journal} {Jpn. J. Appl. Phys}\ }\textbf {\bibinfo {volume} {55}},\ \bibinfo {pages} {08P} (\bibinfo {year} {2016})}\BibitemShut {NoStop}%
\bibitem [{\citenamefont {O'Brien}\ \emph {et~al.}(2000)\citenamefont {O'Brien}, \citenamefont {Koitzsch},\ and\ \citenamefont {Nemanich}}]{OBrien2000}%
  \BibitemOpen
  \bibfield  {author} {\bibinfo {author} {\bibfnamefont {M.~L.}\ \bibnamefont {O'Brien}}, \bibinfo {author} {\bibfnamefont {C.}~\bibnamefont {Koitzsch}},\ and\ \bibinfo {author} {\bibfnamefont {R.~J.}\ \bibnamefont {Nemanich}},\ }\href@noop {} {\bibfield  {journal} {\bibinfo  {journal} {J. Vac. Sci. Technol. B Microelectron. Nanometer Struct. Process. Meas. Phenom.}\ }\textbf {\bibinfo {volume} {18}},\ \bibinfo {pages} {1776} (\bibinfo {year} {2000})}\BibitemShut {NoStop}%
\bibitem [{\citenamefont {Fulton}\ \emph {et~al.}(2006)\citenamefont {Fulton}, \citenamefont {Lucovsky},\ and\ \citenamefont {Nemanich}}]{Fulton2006}%
  \BibitemOpen
  \bibfield  {author} {\bibinfo {author} {\bibfnamefont {C.~C.}\ \bibnamefont {Fulton}}, \bibinfo {author} {\bibfnamefont {G.}~\bibnamefont {Lucovsky}},\ and\ \bibinfo {author} {\bibfnamefont {R.~J.}\ \bibnamefont {Nemanich}},\ }\href@noop {} {\bibfield  {journal} {\bibinfo  {journal} {J. Appl. Phys.}\ }\textbf {\bibinfo {volume} {99}},\ \bibinfo {pages} {063708} (\bibinfo {year} {2006})}\BibitemShut {NoStop}%
\bibitem [{\citenamefont {Wager}(2017)}]{Wager2017}%
  \BibitemOpen
  \bibfield  {author} {\bibinfo {author} {\bibfnamefont {J.~F.}\ \bibnamefont {Wager}},\ }\href@noop {} {\bibfield  {journal} {\bibinfo  {journal} {AIP Adv.}\ }\textbf {\bibinfo {volume} {7}},\ \bibinfo {pages} {125321} (\bibinfo {year} {2017})}\BibitemShut {NoStop}%
\bibitem [{\citenamefont {Singh}\ and\ \citenamefont {Shimakawa}(2003)}]{Singh2003}%
  \BibitemOpen
  \bibfield  {author} {\bibinfo {author} {\bibfnamefont {J.}~\bibnamefont {Singh}}\ and\ \bibinfo {author} {\bibfnamefont {K.}~\bibnamefont {Shimakawa}},\ }\href@noop {} {\emph {\bibinfo {title} {Advances in Amorphous Semiconductors}}}\ (\bibinfo  {publisher} {``Taylor \& Francis"},\ \bibinfo {year} {2003})\BibitemShut {NoStop}%
\bibitem [{\citenamefont {Frenkel}(1938)}]{Frenkel1938}%
  \BibitemOpen
  \bibfield  {author} {\bibinfo {author} {\bibfnamefont {J.}~\bibnamefont {Frenkel}},\ }\href@noop {} {\bibfield  {journal} {\bibinfo  {journal} {Phys. Rev.}\ }\textbf {\bibinfo {volume} {54}},\ \bibinfo {pages} {647} (\bibinfo {year} {1938})}\BibitemShut {NoStop}%
\end{thebibliography}
%

%

\end{document}


\setcounter{figure}{0}
\renewcommand{\figurename}{SUPPL.~FIG.}
\renewcommand\thefigure{\arabic{figure}}
\setcounter{table}{0}
\renewcommand{\tablename}{SUPPL.~TABLE}
\renewcommand\thetable{\arabic{table}}
\renewcommand{\theequation}{S\arabic{equation}}
\setcounter{equation}{0}  

\preprint{APS/123-QED}

\title{Supplementary Information: Using optical tweezers to simultaneously trap, charge and measure the charge of a microparticle in air}

\author{Andrea Stoellner}
\affiliation{Institute of Science and Technology Austria, Am Campus 1, 3400 Klosterneuburg, Austria}
\author{Isaac C.D. Lenton}
\affiliation{Institute of Science and Technology Austria, Am Campus 1, 3400 Klosterneuburg, Austria}
\author{Artem G. Volosniev}
\affiliation{Center for Complex Quantum Systems, Department of Physics and Astronomy, Aarhus University, Ny Munkegade 120, DK-8000 Aarhus C, Denmark}
\author{James Millen}
\affiliation{King's College London, Strand Campus, London WC2R 2LS, United Kingdom}
\author{Renjiro Shibuya}
\affiliation{Chiba University, 1-33, Yayoi-cho, Inage-ku, Chiba 263-8522 Japan}
\author{Hisao Ishii}
\affiliation{Chiba University, 1-33, Yayoi-cho, Inage-ku, Chiba 263-8522 Japan}
\author{Dmytro Rak}
\affiliation{Institute of Science and Technology Austria, Am Campus 1, 3400 Klosterneuburg, Austria}
\affiliation{Institute of Experimental Physics, Slovak Academy of Sciences, Watsonova 47, 040 01 Košice, Slovakia}
\author{Zhanybek Alpichshev}
\affiliation{Institute of Science and Technology Austria, Am Campus 1, 3400 Klosterneuburg, Austria}
\author{Grégory David}
\affiliation{ETH Zürich, Rämistrasse 101, 8092 Zürich, Switzerland}
\author{Ruth Signorell}
\affiliation{ETH Zürich, Rämistrasse 101, 8092 Zürich, Switzerland}
\author{Caroline Muller}
\affiliation{Institute of Science and Technology Austria, Am Campus 1, 3400 Klosterneuburg, Austria}
\author{Scott Waitukaitis}
\email{scott.waitukaitis@ista.ac.at}
\affiliation{Institute of Science and Technology Austria, Am Campus 1, 3400 Klosterneuburg, Austria}

\date{\today}


\maketitle

\section{Sample preparation}
We prepare our sample solution using a monodisperse powder of uncoated, amorphous SiO\textsubscript{2} spheres with a median particle diameter of 0.69 \textmu m and a density of 2.0 g/cm\textsuperscript{3}. (Cospheric SiO2MS-2.0). We suspend approximately 1 wt\% dry particles in Milli-Q\textsuperscript{\textregistered}, working with a vortex mixer and gentle sonication to aid suspension and break up particle clusters. We then aerosolize the solution using a commercial nebulizer (PARI BOY Junior), creating droplets with a mass median aerodynamic diameter of 2.8 \textmu m. The aerosol is sprayed into the chamber from the top for trapping, where the water quickly evaporates, leaving only the silica particle. Once a single particle is trapped, excess particles in the chamber are removed using a vacuum pump. In an alternative preparation protocol we plasma cleaned the dry particles before suspension to probe the role of surface impurities or atmospheric adsorbates – however, there was no difference in charging found between the plasma cleaned and the non-plasma cleaned microparticles. 

\section{Charge measurement principle}
We measure the charge on our particle following the method of Ref.~\cite{Ricci2019}. The particle acts as a driven harmonic oscillator and obeys the equation of motion
\begin{equation}
    m\ddot{x} + m\Gamma\dot{x} + kx = \sqrt{2 k_B T m \Gamma} \eta(t) + Q E(t),
    \label{eq:equation of motion}
\end{equation}
where $m$ is the particle mass, $\Gamma$ is the damping rate and $k = m \Omega_0^2$ is the optical trap stiffness ($\Omega_0$ is the mechanical eigenfrequency). The first term on the right side of the equation describes the thermal driving of the particle by Brownian motion, where $\eta(t)$ represents white noise. The second term describes the electric driving we apply, where the electric field is of the form $E(t) = E_0 cos(\omega_{dr}t)$. Solving for the particle charge, $Q$, and calculating the power spectral density (PSD) leads to the form

\begin{equation}
    S(\omega) = S_{th}(\omega) + S_{el}(\omega) = \frac{4 k_B T \Gamma}{m[(\omega^2 - \Omega_0^2)^2 + \Gamma^2 \omega^2]} + \frac{Q^2 E_0^2 \tau \mathrm{sinc}^2[\frac{\tau}{2} (\omega - \omega_{dr})]}{2 m^2 [(\omega^2 - \Omega_0^2)^2 + \Gamma^2 \omega^2]},
    \label{eq:PSD 1}
\end{equation}
where $\tau$ is the time over which the particle is being observed. The experimentally measured PSD $S_{exp}(\omega)$ is related to $S(\omega)$ by the calibration factor $c_{cal}$ so that $S_{exp}(\omega) = c_{cal}^2 S(\omega)$. For the charge measurement, however, $c_{cal}$ cancels out and we can use the above expression. The charge of the particle can be calculated by comparing the electrically driven ($S_{el}$) and thermally driven ($S_{th}$) fractions of the PSD at the driving frequency of the electric field $\omega_{dr} = 2 \pi f_{dr}$

\begin{equation}
    \frac{S_{el}(\omega_{dr})}{S_{th}(\omega_{dr})} = \frac{Q^2 E_0^2 \tau }{8 m k_B T \Gamma}.
    \label{eq:charge equation}
\end{equation}
From there we can easily find the particle charge Q
\begin{equation}
    Q = \sqrt\frac{8 k_B T \gamma S_{el}(w_{dr})} {E_0^2 \tau S_{th}(w_{dr})},
    \label{eq:charge equation}
\end{equation}
where we used $\gamma = m\Gamma$. Experimentally, we measure $S_{el}(\omega_{dr}) = S(\omega_{dr}) - S_{th}(\omega_{dr})$, where $S(\omega_{dr})$ is the value of the PSD at the charge peak (see Fig. 2(a) in the main manuscript). To extract the sign of the particle's charge, we additionally monitor the phase of the position signal relative to the electric field. 

\section{Full charging curves and discharges at long times}

\begin{figure}[H]
\centering
\includegraphics{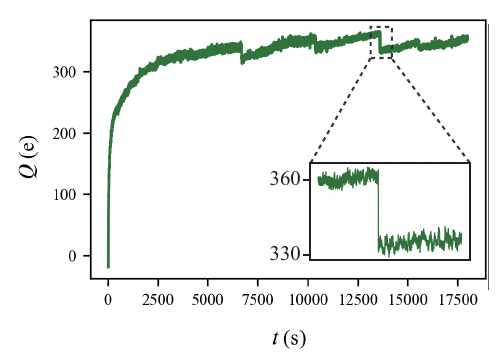}
\caption{\label{fig:SI discharges} Long-time charging curve taken at 100 mW, showing (a) that the primary charging mechanism continues beyond what we show in the main draft and (b) that eventually the particle becomes so charged that spontaneous discharges occur due to electrical breakdown. The inset shows a discharge with a magnitude of around 30e. 
}
\end{figure}

The curves in Fig.~3(a) of the main manuscript are limited to $\sim$2000 s in time to better illustrate the variable charging rate at different laser powers. However, the charging behavior for each power was measured for much longer during the experiment. Suppl.~Fig.~\ref{fig:SI discharges} shows an example curve with a duration of approximately five hours taken at 100 mW. This illustrates that the primary mechanism causes the charge to continue growing even beyond what we show in the main text. It also shows, however, that a particle can become so charged such that spontaneous discharges occur, visible in Suppl.~Fig.~\ref{fig:SI discharges} as sudden drops in charge (\textit{e.g.}, in the inset). These are highly likely to be caused due to electrostatic breakdown of the air around the particle. Assuming discharges will occur when the electric field at the particle surface exceeds the dielectric strength of air ($E_{bd}\approx 3\times10^6$ V/m), then the corresponding charge we expect a particle to have is
\begin{equation}
Q_{max} = E_{bd}4\pi\epsilon_0 r^2\approx 250 \, \,|e|.
\end{equation}
This limit is in good agreement with what we observe, and effectively caps the maximum charge a particle can attain. The discharges can range from just a few up to multiple tens of elementary charges. The physics behind these ``micro-discharges'' is highly relevant to topics like cloud electrification, where it is suspected that such an event may be able to trigger macroscopic sparks in sub-critical fields (see \textit{e.g.}, \cite{Petersen2014}). This is one of the primary future directions we will pursue with our system.



\section{Estimation of particle temperature}
We estimate the temperature of the trapped silica sphere by assuming equilibrium between the optical power absorbed and the heat dissipated through conduction to the surrounding air. For a given laser power $P$, the temperature $T_p$ of the sphere is given by

\begin{equation}
    T_p = \frac{P \alpha d}{4 \pi r \kappa} + T_0,
    \label{eq:temperature}
\end{equation}
where $\alpha$ is the absorption coefficient of silica at our wavelength, $d$ is the length scale over which the light is absorbed (in this case the particle diameter), $r$ is the particle radius, $\kappa$ is the heat conductivity of air and $T_0$ is the ambient temperature. For our maximum applied laser power $P = 200$ mW and $\alpha \sim 2 \times 10^{-7} \mu m^{-1}$ (Supplementary Information \cite{Bateman2014}), we find that the resulting temperature increase is $\sim0.25$ K--—indicating that the particle does not heat significantly under our trapping conditions, even for the conservative assumption that all optical power delivered to the trap is incident on the particle.

\section{Photoelectron yield spectroscopy (PYS)}

\textbf{Sample Preparation} We prepare our PYS samples using a monodisperse powder of uncoated, amorphous SiO\textsubscript{2} spheres with a median particle diameter of \SI{0.69}{\micro\meter} and a density of 2.0 g/cm\textsuperscript{3} (Cospheric, SiO2MS-2.0) in bulk powder form. The particles are dispersed in toluene at \SI{2}{wt\percent}, stirred at \SI{800}{rpm} for \SI{1}{hour}, sonicated for \SI{20}{minutes}, and stirred again for \SI{30}{minutes}. A \SI{50}{\micro\liter} aliquot is drop-cast onto a p-type (100) silicon substrate, pre-cleaned with acetone and isopropanol, and dried under ambient conditions. A second \SI{50}{\micro\liter} aliquot is then applied to complete the film. In addition to PYS, for the constant final state photoelectron yield spectroscopy (CFS-YS) measurement, we use a thermally oxidized p-type (100) silicon wafer with a \SI{100}{\nano\meter} SiO\textsubscript{2} layer, cleaned in the same way.

\vspace{0.2em}

\textbf{PYS and CFS-YS Method} In PYS, the emitted photoelectrons were detected using an electron multiplier (Channeltron CEM 4821G), while a voltage of \SI{-10}{\volt} was applied to the sample. The entrance electrode of the electron multiplier was biased at \SI{100}{\volt} to attract the emitted photoelectrons. A deuterium lamp (Hamamatsu Photonics: L1835, \SI{150}{\watt}) and a xenon lamp (Ushio: UXL-500D, \SI{500}{\watt}) were used as light sources (\( h\nu = \SI{7.7}{eV} - \SI{3}{eV} \)), and the emitted light was monochromatized using a zero-dispersion double monochromator (Bunkoukeiki: BIP-M25-GTM). A photomultiplier tube (Hamamatsu Photonics: R376, R6836) estimated the photon flux to be approximately \(10^7 - 10^{12}~\text{cm}^{-2}\text{s}^{-1}\).

CFS-YS was performed by irradiating a sample with light with a photon energy of \SIrange{1.5}{7.7}{eV}. Electrons emitted from the sample surface were collected using an electrostatic hemispherical analyzer (PSP: RESOLVE120). The kinetic energy of measured photoelectrons was set to be \SI{0.54}{eV}. The photoelectron emission properties were determined in a measurement chamber maintained at \(2 \times 10^{-6}~\text{Pa}\). Also, charging treatment to inject electrons was performed by applying a positive bias to the sample holder while operating the ion gauge, using its filament as the electron source.

\begin{figure}[H]
\centering
\includegraphics[width=1\textwidth]{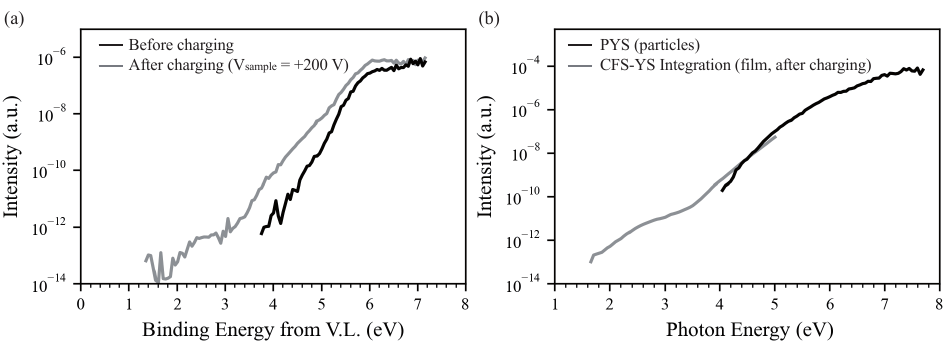}
\caption{\label{fig:Hisao} (a) CFS-YS spectrum as a function of binding energy from vacuum level (V.L.) of thermally grown SiO\textsubscript{2} before and after charging treatment with a \SI{200}{\volt} sample bias. (b) PYS spectrum of silica particles (see Fig. 4(c) in the main manuscript) and rescaled, integrated CFS-YS spectrum of thermally grown SiO\textsubscript{2} charged with a \SI{200}{\volt} sample bias.}
\end{figure}

\textbf{Results} Suppl. Fig. \ref{fig:Hisao} shows the CFS-YS spectra before and after charging by electron beam (with \SI{200}{\volt} sample bias), reflecting the density of states (DOS) of the SiO\textsubscript{2} film. Before charging, the spectrum onset appears around \SI{4}{eV}, similar to silica particles, indicating the presence of electronic states up to that energy. After charging negatively, the onset shifts to approximately \SI{2}{eV}, and spectral intensity increases across the full energy range, suggesting that injected electrons filled previously unoccupied states below \SI{4}{eV}. The observed shallow states are due to negative, not positive, charging, but it is clear that there is non-negligible DOS below \SI{4}{eV} in the SiO\textsubscript{2} film. In the optical tweezers experiment (Fig. 4(c) in the main manuscript), photoemission was observed from energy levels shallower than \SI{4}{eV}. The states above \SI{4}{eV} are unoccupied in the neutral condition but can be partially occupied even in positively charged conditions under laser irradiation, because one phonon absorption of SiO\textsubscript{2} can induce many electron excitations to these states. The grey line in Suppl. Fig. \ref{fig:Hisao} shows the integrated CFS-YS spectrum after charging, obtained by integrating the spectral intensity from the onset up to each energy value and plotting the cumulative values as a function of energy. This analytical approach gives the simulation of PYS, and it revealed that an exponentially decaying structure exists below \SI{4}{eV} after electron injection. Comparison with the PYS spectrum of the particles (black line) shows a reasonable overlap and suggests the presence of an exponentially decaying structure, which was not detectable by PYS alone. In PYS, the electron excitation to the shallow unoccupied state above \SI{4}{eV} is not observed because of the low photon density, but it is expected in the optical tweezers experiment due to the laser beam.

\section{Two-photon assisted thermal emission model}

In the thermal emission model, the two-photon process promotes electrons from deeper levels ($> 4.66eV$) to an excited state, \textit{e.g.} in the conduction band. From there, a small fraction of the excited electrons escapes thermally, therefore leading to the observed charging. Mathematically, we consider first the rate equation for the number of electrons in the excited state, $N_e$, given by 
\begin{equation}
    \frac{dN_e}{dt} = k_1I^2 - k_2 N_e - K(Q)N_e,
    \label{eq:dN/dt}
\end{equation}
where the first term is the initial two-photon process ($k_1$ being a rate constant), the second term corresponds to spontaneous decay back to the initial state ($k_2$ being a decay constant), and the third term captures the thermal ejection from the particle we are ultimately interested in ($K(Q)$ being a coefficient that depends on the net particle charge, $Q$). Assuming the third term is small compared to the first two, the quasi-steady state population in the excited state is
\begin{equation}
    N_e = \frac{k_1}{k_2}I^2.
    \label{eq:N}
\end{equation}
From this reservoir of intermediate-state electrons, the rate of those that escape completely is simply given by the third term. Substituting Eq.~\ref{eq:N} for $N_e$, we find
\begin{equation}
    \frac{dQ}{dt} = \frac{k_1}{k_2}I^2 K(Q),
    \label{eq:dQ/dt}
\end{equation}
where we keep our convention that $Q$ is in units of electrons. Given our assumption that escape from the excited state is thermal, $K(Q)$ should be given by the Boltzmann factor, \textit{i.e.}~

\begin{equation}
    K(Q) \propto e^{-b_t \frac{Qe}{4\pi\epsilon_{\scalebox{0.40}0} r k_{\scalebox{0.40}B} T}}.
    \label{eq:c/gamma f(N)}
\end{equation}
where $k_BT$ is the Boltzmann energy, $\epsilon_0$ is the vacuum permittivity, $r$ is the radius of the particle and $b_t$ is a constant of order one. Here, we have absorbed the contribution from the initial energy deficit defined in the main manuscript ($\delta=\epsilon-2h\nu$) into the proportionality. Inserting Eq.~\ref{eq:c/gamma f(N)} into Eq.~\ref{eq:dQ/dt} and absorbing $k_1$ and $k_2$ into a single constant, $a_t$, yields 

\begin{equation}
    \frac{dQ}{dt} = a_t I^2 e^{-b_t\frac{Qe}{4\pi\epsilon_{\scalebox{0.40}0} r k_{\scalebox{0.40}B} T}}.
    \label{eq:dQ/dt full}
\end{equation} 
This differential equation is easily solved. Enforcing the initial condition $Q(t=0)=0$ yields

\begin{equation}
    Q(t) = \frac{4 \pi \epsilon_0 r k_B T}{b_t e} \ln \bigg{(}\frac{e}{4 \pi \epsilon_0 r k_B T}a_t b_t I^2 t + 1 \bigg{)}.
\label{eq:Q(t)}
\end{equation}
In practice, we define a new constants $B = \frac{b_t}{k_B T}$ and redefine $A = a_t$, which leads to the fitting equation

\begin{equation}
    Q(t) = \frac{4 \pi \epsilon_0 r}{B e} \ln \bigg{(}\frac{e}{4 \pi \epsilon_0 r}A B I^2 t + 1 \bigg{)},
\label{eq:Q(t) thermal}
\end{equation}
where $Q(t)$ is in number of elementary charges. All curves, regardless of power, fit to the same two constants, $A$ and $B$. We find the best fitted values are $A = 6.4\times10^{-2}$ e m$^4$ s$^{-1}$ GW$^{-2}$ and $B = 4.9$ eV$^{-1}$. For $k_B T = 0.026$ eV the original parameters from Eq.~\ref{eq:Q(t)} are $a_t = 6.4\times10^{-2}$ e m$^4$ s$^{-1}$ GW$^{-2}$ and $b_t = 0.13$. 

\section{Two-photon direct emission model}
The direct emission model assumes that the charge we measure originates from electrons ejected directly from shallow electronic states. For an uncharged particle the available ionization energy is the two-photon energy $2h\nu = 4.66$~eV. However, as the particle becomes more and more charged, the electrostatic potential energy acts as an additional energetic barrier that the electrons need to overcome using the energy $2h\nu$. By defining the ``effective'' ionization energy 

\begin{equation}
    \epsilon(Q) = 2h\nu - \frac{Qe}{4 \pi \epsilon_0 r}
    \label{eq:ionization energy}
\end{equation}
and plotting it against the charging rate $\frac{dQ}{dt}$ (green curve in Fig. 4(c) in the main draft), we can find an upper bound for the energy depth at which the electron of each specific charging event was ejected. As shown in the main manuscript we find electronic states as low as 3 eV (and for longer measurement times potentially even lower) and that the number of occupied electronic states grows exponentially with energy. We therefore define the number of electronic states available at a given effective ionization energy

\begin{equation}
    N_{available}(Q) = c e^{b_d \epsilon(Q)},
    \label{eq:DOS 1}
\end{equation}
where $c$ and $b_d$ are constants. The two-photon process can be seen as an attempt rate of electron emission $\propto I^2$, liberating electrons from the available electron states $N_{available}(Q)$. Combining the two and inserting Eq.~\ref{eq:ionization energy} into Eq.~\ref{eq:DOS 1} we find the charging rate is given by

\begin{equation}
    \frac{dQ}{dt} = a_d I^2 e^{b_d (2h\nu - \frac{Qe}{4 \pi \epsilon_0 r})} 
    \label{eq: DOS4},
\end{equation}
where we have absorbed $c$ into the two-photon rate constant $a_d$,  $\epsilon_0$ is the vacuum permittivity and $r$ is the radius of the particle. For $Q(t=0)=0$, the solution is

\begin{equation}
    Q(t) = \frac{4 \pi \epsilon_0 r}{b_d e} \ln \bigg{(}\frac{e}{4 \pi \epsilon_0 r} a_d e^{-b_d 2 h \nu} b_d I^2 t + 1 \bigg{)},
\label{eq:DOS Q(t)}
\end{equation}
To bring the equation into the same form as Eq. \ref{eq:Q(t) thermal} we define a new constant $A = a_d e^{-b_d 2 h \nu}$ and rename $B = b_d$, which leads to the same fitting equation

\begin{equation}
    Q(t) = \frac{4 \pi \epsilon_0 r}{B e} \ln \bigg{(}\frac{e}{4 \pi \epsilon_0 r} A B I^2 t + 1 \bigg{)}.
\label{eq:DOS5}
\end{equation}
As before we find $A = 6.4\times10^{-2}$ e m$^4$ s$^{-1}$ GW$^{-2}$ and $B = 4.9$ eV$^{-1}$. The original parameters from Eq. \ref{eq:DOS Q(t)} are $a_d = 5.3\times10^{8}$ e m$^4$ s$^{-1}$ GW$^{-2}$ and $b_d = 4.9$ eV$^{-1}$.

\bibliographystyle{apsrev4-2}